# Using Positional Sequence Patterns to Estimate the Selectivity of SQL LIKE Queries

Mehmet Aytimur
Department of Computer Science
Istanbul Sehir University
mehmetaytimur@std.sehir.edu.tr

Ali Cakmak[1]
Department of Computer Science
Istanbul Sehir University
alicakmak@sehir.edu.tr

## Abstract

*With the dramatic increase in the amount of the text-based data which commonly contains misspellings and other errors, querying such data with flexible search patterns becomes more and more commonplace. Relational databases support the LIKE operator to allow searching with a particular wildcard predicate (e.g., LIKE 'Sub%', which matches all strings starting with 'Sub'). Due to the large size of text data, executing such queries in the most optimal way is quite critical for database performance. While building the most efficient execution plan for a LIKE query, the query optimizer requires the selectivity estimate for the flexible pattern-based query predicate. Recently, SPH algorithm is proposed which employs a sequence pattern-based histogram structure to estimate the selectivity of LIKE queries. A drawback of the SPH approach is that it often overestimates the selectivity of queries. In order to alleviate the overestimation problem, in this paper, we propose a novel sequence pattern type, called positional sequence patterns. The proposed patterns differentiate between sequence item pairs that appear next to each other in all pattern occurrences from those that may have other items between them. Besides, we employ redundant pattern elimination based on pattern information content during histogram construction. Finally, we propose a partitioning-based matching scheme during the selectivity estimation. The experimental results on a real dataset from DBLP show that the proposed approach outperforms the state of the art by around 20% improvement in error rates.*

## 1. Introduction

Query optimization plays a critical role on the performance of database management systems. In a database management system, the optimizer goes over all or a subset of possible execution plans, and determines the one with the lowest cost. A critical input to the optimizer while deciding on the most optimal query plan is the selectivity of each predicate in a query. The selectivity of a predicate *p* indicates the fraction of database rows that would be retained after applying *p* as a filter on the rows. Predicate selectivity is usually estimated based on statistics gathered from the database and stored in the database catalog before query time (i.e., offline). The accuracy of such predictions are vital for the optimizer to come up with an efficient execution plan. With number-typed data, the query optimizers usually come up with sufficiently accurate predictions owing to vast past research. On the other hand, for predicates on text-based data, it is still a challenge for query optimizers to come up with reliably accurate selectivity estimates. The main reason for this disparity is that predicates on text-based data often contain flexible patterns that may match a wide array of rows. In contrast, predicates on numeric data are strict and well-formed. The particular need for flexible predicates on text data results from the fact that textual data is frequently not clean with many misspellings and typographical errors. Especially with the explosion of the internet resources, the "dirtiness" of textual data is even more evident. Furthermore, due to the nature of the text data, shared character sequences (e.g., prefixes, suffixes, etc.) often determine a particular class of data rows that database users may often be interested in. SQL provides the LIKE operator to allow formulating wildcard predicates on text data. For instance, the predicate *name LIKE 'Luck%'* returns all the names that start with 'Luck'. Accurately predicting the selectivity of such flexible predicates on text data is still an open research problem.

There are several works (Jagadish, Ng, & Srivastava, 1999; Jagadish, Kapitskaia, Ng, & Srivastava, 2000; Krishnan, Vitter, & Iyer, 1996; Lee, Ng, & Shim, 2009) in the literature which study the selectivity estimation of the wildcard predicates. The majority of past studies often assume that the predicates are in the form of *%w%* where *w* represents more than one characters, and they perform poorly (Lee et al., 2009) for more generic queries that are in the form of $\%s_1\%s_2\%...\%s_n\%$ where $s_i$ represents one or more characters. Besides, the main characteristic of these techniques is that they process all or a large set of row ids in the database during selectivity prediction. Processing such a large data for each query during optimization time (i.e., online) suffers from memory and execution time overhead.

[1] Corresponding Author. Address: Orhantepe Mh. Turgut Ozal Blvd. No:21, Kartal 34865, Istanbul, Turkey

We recently proposed an algorithm, called SPH, to predict the selectivity of LIKE query predicates (Aytimur & Çakmak, 2018). In SPH approach, first, frequent sequence patterns are computed from the database before query time. Then, a histogram structure is built out of the discovered patterns. In order to estimate the selectivity of a given LIKE predicate, the precomputed histogram is exploited during query optimization time. SPH works reasonably well in many settings, it often overestimates the selectivity of LIKE predicates that are in the form of $\%s_1s_2\%s_3\%s_4s_5\%...\%s_n\%$ where $s_i$ represents a character. We give an example.

**Example:** Let "*%A%B%C%D%A%*" be a bucket endpoint value in histogram with frequency 10. If LIKE predicate p is "*%AB%C%DA%*", then SPH estimates the number of rows in the result set of this query as 10 based on the above histogram endpoint pattern. However, the true frequency of "*%AB%C%DA%*" such a predicate may often be significantly less than 10.

We observed three factors that contribute to the overestimation in SPH as follows:

(i) The employed patterns are too generic, which makes SPH fail to differentiate the consecutively placed characters from those that are not strictly consecutive and may have some other characters between them.

(ii) Histogram endpoints are often occupied by patterns that have high frequency and are almost completely subsuming one another with little difference in the information that they provide. This greatly narrows down the coverage and diversity of SPH histograms, and leads to the elimination of many other potentially useful patterns from consideration as endpoints during histogram construction.

(iii) SPH attempts to fully match a query predicate pattern to histogram endpoints, and ignore partial matches. This prevents SPH from taking advantage of the constructed histogram, and contributes to its overestimation rate.

In order to address the above issues, in this paper, we extend SPH and propose using a new type of sequence pattern form that discriminates between consecutively appearing characters and those that may not. Such patterns are not supported by the existing sequence pattern mining algorithms (Pasquier, Bastide, Taouil, & Lakhal, 1999; Wang, Han, & Pei, 2003; Wang & Han, 2004; Yan, Han, & Afshar, 2003; Zaki & Hsiao, 2002) in the literature. Hence, we propose a new sequence mining algorithm to compute patterns of the form, $\%a_1\%a_2\%...\%a_n\%$ where each $a_i$ represents one or more characters. Note that in standard sequence pattern form, each $a_i$ represents a single character. We call these new kinds of patterns *positional sequence patterns*. Like SPH, we build histograms out of the computed patterns, and these histograms are later used to estimate the selectivity of LIKE predicates. However, we eliminate redundant patterns that harm the diversity and coverage of the constructed histogram. Besides, we integrate a partial matching framework during selectivity estimation. We call the proposed approach in this paper *P-SPH* (*P* for positional sequence patterns).

We use the DBLP dataset in order to extensively evaluate our methods. Our results show that P-SPH decreases the error rate of selectivity estimations up to 20% in comparison to the state of the art.

***Contributions*:** Our primary contributions in this paper are as follows:

- In order to address the first of the above issues, we propose a new type of sequence pattern (i.e., positional sequence patterns) that carries more information than the standard sequence patterns which do not specify whether there is any character between two consecutive items in the sequence or not. That is, if two characters always appear at consecutive positions in all appearances of a pattern, such information is also marked in the new type of patterns as well.
- Since positional sequence patterns are not considered by the existing sequence mining literature, we propose an algorithm to compute this new pattern type by extending a standard sequence pattern mining algorithm (Wang & Han, 2004).
- As for the second issue, during histogram construction out of the mined patterns, we introduce *information content-based* elimination of some patterns that highly overlap with other histogram endpoint patterns.
- In order to address the third issue, we propose a slider-based partial pattern matching scheme during the selectivity estimation of a LIKE query predicate pattern based on a histogram built in the previous phase.
- We assess the proposed algorithms comparatively in different aspects on real datasets, and show that it outperforms the state of the art approaches.

The rest of this paper is organized as follows. In section 2, we discuss the related work. Section 3 describes the computation of positional sequence patterns. In section 4, we present building histograms out of positional sequence patterns. Section 5 describes the selectivity estimation steps. In section 6, we present a detailed experimental evaluation of the proposed methods, and Section 7 concludes.

## 2. Related Work

Sequential pattern mining was first proposed in (Agrawal & Srikant, 1995 ). Since then, a number of algorithms have been proposed to compute sequential patterns in a dataset. SPADE (Zaki, 2001), PrefixSpan (Pei et al., 2004), and SPAM (Ayres, Gehrke, Yiu, & Flannick, 2002) are some of the well-known sequence mining approaches. SPADE is based on vertical id-list database format, and uses a lattice-theoretic approach to decompose the original search space into smaller spaces. PrefixSpan employs a horizontal format, and uses the pattern-growth method. SPAM adopts a vertical bitmap representation, and mines the longest sequential patterns. Recently, VMSP (Viger, Wu, Gomariz, and Tseng, 2014) is proposed. VMSP is based on vertical id-list format, and mines not all sequential patterns, but only the “maximal” sequential patterns. Wang et al. (Wang & Han, 2004) et. propose BIDE, and similarly, rather than mining all frequent sequences, it mines only “closed” patterns.

The selectivity estimation of wildcard predicates has been studied extensively in the previous studies. Jin and Li (2015) propose an algorithm, called SEPIA. They employ a kind of frequency table in order to keep some summary data. To, Chiang, and Shahabi (2013) propose information entropy-based histograms in order to estimate the selectivity of a LIKE predicate. They develop three algorithms (ME, MSE and, MB) which are based on the information entropy, and they estimate the selectivity by using information entropy-based histograms. Jagadish et al. (1998) exploit optimal histograms to estimate the selectivity. They propose an algorithm which finds the optimal bucket boundaries. Poosala, Haas, Ioannidis, and Shekita (1996) offer equi-width and equi-height histogram structures. Muralikrishna and DeWitt (1988) propose an algorithm to build multi-dimensional histograms to estimate the selectivity of multi-dimensional queries.

Krishnan et al. (1996) employ suffix trees in order to structure the textual data, and estimate the selectivity for the substrings in a wildcard predicate. They propose an algorithm called KVI, and it assumes the independence of substrings. Suffix tree approach is used in many other studies, as well. Jagadish, Ng, and, Srivastava (1999) propose MO algorithm in order overcome the accuracy problem due to KVI’s independence assumption. The MO algorithm is based on Markov assumption. Chaudhuri, Ganti, and Gravano (2004) observe that MO often underestimates selectivities, and they introduce a new algorithm called CRT which is based on the Short Identifying Substring (SIS) assumption. Lee et al. (2009) exploit both suffix tree structure and MO as tools in their approach, and propose two new algorithms, MOF and LBS. They use the edit distance in order to find all base substrings of the given query predicate pattern. They estimate the selectivity by using minimal base substrings and information stored in an N-gram table. MOF is essentially an extended version of the authors’ earlier work (Lee, Ng, and, Shim, 2007). MOF is further extended into LBS, which assigns a signature to each minimal base substring, and keeps them in an N-gram table with their database frequencies. As for the minimal base substrings which are not stored in the N-gram table, LBS employ MO and suffix trees to estimate the frequency. Moreover, Kim, Woo, Park, and Shim (2010) use a similar approach for the selectivity estimation. Their method is based on inverted-gram indices. They employ signature generated by random permutation for each substring. Both techniques proposed by Lee et al. and Kim et al. are grounded on similar techniques, and their reported results are almost the same. Besides, Mazeika, Koudas and, Srivastava (2007) propose VSol to estimate the selectivity of the approximate string queries. They use edit distance and q-grams for estimation. In order to access the q-grams, and estimate the selectivity for a query, they use a hash index. VSol’s approach is very similar to LBS (Lee et al.,2009), which is discussed above.

Approximate string matching is also popular in spatial databases. As an example, Yao, Li, Hadjieleftheriou, and Hou (2010) use MHR tree to efficiently answer approximate string matching queries in large spatial databases. Their technique is based on the min-wise signature and the linear hashing technique, which are also used in LBS approach.

Recently, Aytimur and Çakmak (2018) propose a novel approach, called SPH. SPH first mines all frequent closed sequence patterns by using an existing sequence mining algorithm (Lee et al. 2004). Then, it builds a histogram structure from mined patterns. Histograms store the sequence patterns as their endpoint values, and their corresponding frequencies as endpoint counts. During the selectivity estimation, all buckets of the histogram are visited, and based on the nature of matching between the histogram bucket endpoints and the predicate, it returns an estimated selectivity value. SPH is currently the state of the art approach to estimate the selectivities of LIKE query predicates.

Our approach, in this study, extends SPH in the following aspects: (i) we compute and employ a new type of sequential pattern which accommodates strictly consecutively appearing character pairs as part of the pattern, (ii) during histogram construction, rather than using all computed patterns blindly, we carefully filter patterns that sufficiently differ from others based on information theoretic measures, (iii) SPH only considers full pattern matching, while our approach in this work also features partial pattern matching in its selectivity estimation step with a slider-based search.

## 3. Positional Sequence Patterns

Sequence mining, in general, aims to find the statistically relevant sequence patterns in a given database. Since its introduction in 90s, it has been used in a wide range of applications such as analyzing the DNA and RNA sequences to figure out coding regions, discovering customer shopping behaviors, telephone calling patterns in call centers, and finding user click patterns in web click streams. Standard (regular) sequence patterns, and the problem of mining such patterns may be formally defined as follows.

**Def'n** *(Proper Sequence Containment)***:** Given two sequences $S = s_1s_2 \ldots s_n$ and $Q = q_1q_2 \ldots q_k$ where each $s_i$ ($1 \leq i \leq n$) and $q_j$ ($1 \leq j \leq k$) are characters from an alphabet, and i, j indicate the position of characters in S and Q, respectively, let Q[i] denote the character at position i in Q. Then, S is *properly contained* in Q, if there exists a set of n positions, $p_1 < p_2 < \ldots < p_n$, such that;

$$(s_i, s_j) \;\; S \text{ and } i < j \quad Q[p_i] = s_i \text{ and } Q[p_j] = s_j \text{ and } j - i \leq p_j - p_i$$

**Example:** Consider S = ABD. S is *properly contained* in $Q_1$ = ACBCD. However, S is *not properly contained* in $Q_2$ = ACDCB.

**Def'n** *(Regular Sequence Pattern):* Given a sequence database *D* and a frequency threshold *minsup*, a sequence *P* is called a *regular sequence pattern*, if the number of rows that *properly contain P* in *D* is equal to or greater than *minsup*.

**Def'n** *(Regular Sequence Pattern Mining Problem)***:** Given a sequence database *D* and a frequency threshold *minsup,* compute the set of all regular sequence patterns.

**Example**: Consider a sequence database shown in Figure 1 that has four rows of sequences with letters from the following alphabet, $\Sigma = \{A, B, C, D, E\}$. Assume that the minimum support threshold is 3. Then, the complete set of regular sequence patterns and their corresponding frequencies are as follows {*A*:4, *AB*:4, *AC*:4, *ACB*:4, *ACBE*:3, *ACE*:4, *B*:4, *BA*:3, *BAB*:3, *BAE*:3, *BC*:3, *BCB*:3, *BCE*:3, *BE*:4, *C*:4, *CC*:3, *CCB*:3, *CCE*:3}

| Sequence id | Sequence |
|---|---|
| 1 | ABCABE |
| 2 | BCACDBE |
| 3 | BACDCEDB |
| 4 | ACECBE |

**Figure 1:** An example sequence database

In this paper, we introduce a new type of sequence patterns, called *positional sequence patterns*. The difference between regular sequence patterns and positional sequence patterns is that the latter distinguishes item pairs that always appear next to each other in the same order in all the occurrences of the pattern from those that may have other items between them in some or all occurrences of the pattern. Therefore, positional sequence patterns are more specific and carry more information than regular sequence patterns. We give an example.

**Example:** Consider a regular sequence pattern R = ACCB. In SQL LIKE syntax, this pattern may be expressed as A%C%C%B. That is, between any characters in R, there may be zero or more other characters in the actual occurrences of the pattern. Now, consider a positional sequence pattern in SQL LIKE syntax, P = AC%CB. P carries the extra information that A is always followed by C with no other characters in between in all occurrences of P. In addition, in at least one occurrence of P, there is another character between the double C characters at the middle. Similarly, the last part of the pattern suggests that there is no character between C and B in all occurrences of P.

We next formally define positional sequence patterns in a similar manner that the regular sequence patterns are defined above.

**Def'n** *(Positional Sequence)*: Given an alphabet $\Sigma$, a sequence $S = s_1 s_2 \ldots s_n$ is called a *positional sequence*, if $\forall s_i \in S \rightarrow s_i \in \Sigma \cup \{\text{'\%'}\}$ where i in $s_i$ indicates the position of a particular character in the sequence ($1 \leq i \leq n$).

**Def'n** *(Positional Sequence Containment)*: Given a positional sequence $S = s_1 s_2 \ldots s_n$ and regular sequence $Q = q_1 q_2 \ldots q_k$, let Q[i], where $1 \leq i \leq n$, denote the character at position i in Q, and the slicing operator S[i:j] specifies the subsequence $s = s_i s_{i+1} \ldots s_{j-1} s_j$ of S. Then, S is *positionally contained* in Q if there exists a set of n positions, $p_1 < p_2 < \ldots < p_n$, such that;

$(s_i, s_j)$ S, $s_i \neq$'%', $s_j \neq$'%' and $i < j$ $\quad$ $Q[p_i] = s_i$ and $Q[p_j] = s_j$ and $j - i \leq p_j - p_i$ if '%' S[i+1:m-1]

$Q[p_i] = s_i$ and $Q[p_j] = s_j$ and $j - i = p_j - p_i$ if '%' S[i+1:m-1]

**Example**: Consider S = A%BD. S is *positionally contained* in $Q_1$ = ACBD. However, S is *not positionally contained* in $Q_2$ = ACBAD.

**Def'n** *(Positional Sequence Pattern):* Given a sequence database D, a frequency threshold minsup, a positional sequence P is called a *positional sequence pattern*, if the number of rows that *positionally contain* P in D is equal to or greater than minsup.

**Def'n** *(Positional Sequence Pattern Mining Problem)***:** Given a sequence database D, and a frequency threshold minsup, compute the set of all positional sequence patterns.

**Example**: Consider the sequence database shown in Figure 1. Assume that the minimum support threshold is 3. Then, the complete set of positional sequence patterns (in SQL LIKE predicate syntax) and their corresponding frequencies are as follows {***AC*:3**, ***AC%B*:3**, ***AC%E*:3**, *A*:4, *A%B*:4, *A%C*:4, ***A%C%BE*:3**, *A%C%B*:4, *A%C%E*:4, *B*:4, *B%A*:3, *B%A%B*:3, *B%A%E*:3, *B%C*:3, *B%C%B*:3, *B%C%E*:3, *B%E*:4, *C*:4, *C%C*:3, *C%C%B*:3, *C%C%E*:3} where the patterns in bold are new additions to the list of patterns that are obtained with regular sequence pattern mining.

We next discuss the computation of positional sequence patterns.

### 3.1 Mining Positional Sequence Patterns

There is already a number of available techniques (Pasquier et al., 1999; Wang, Han, & Pei, 2003; Yan, Han, & Afshar, 2003; Zaki & Hsiao, 2002) proposed in the literature to compute regular sequence patterns. Since positional sequence patterns are a superset of regular sequence patterns, rather than designing a new algorithm, we choose to extend one of the existing regular sequence mining methods to compute positional sequence patterns. Among many others, in this work, we determine to extend BIDE (Wang & Han, 2004) for three reasons: (i) it eliminates the candidate maintenance step of standard pattern mining methods; hence, its memory and running time requirements are lower, (ii) rather than mining all possible patterns, it skips 'redundant' patterns, and mines only a subset of all patterns (i.e., 'closed' patterns) which are not contained in other patterns, and (iii) its source code is publicly available to the researchers; thus, it can be readily modified for extensions.

```
Algorithm 1: BIDE
Input:  An input sequence database SDB, a minimum support threshold min_sup.
Output: The complete set of frequent closed sequences, FCS
1. FCS= Ø;
2. F1=frequent 1-sequences(SDB , min_sup);
3. for (each 1-sequence f1 in F1) do
4.     SDB^f1 =pseudo projected database(SDB)
5. for(each f1 in F1) do
6.     if (!BackScan(f1, SDB^f1))
7.    BEI =backward extension check(f1, SDB^f1);
8.    call bide(SDB^f1, f1, min_sup, BEI, FCS);
9. return FCS;

Algorithm : bide
Input:  a projected sequence database Sp_SDB, a prefix sequence Sp,
 a minimum support threshold min_sup, and the number of backward extension items BEI
Output: The current set of frequent closed sequences, FCS
10. LFI = locally frequent items (Sp_SDB;
```

```
11. FEI = |{z in LFI |z.sup = sup^SDB(S_p) }|
12. if ((BEI not equal FEI)==0)
13.    FCS=FCS U {Sp};
14. for (each I in LFI) do
15.      S_p^i <S_p,i>;
16.      SDB^S_pi pseudo projected database (S_p_SDB, S_p^i);
17. for (each I in LFI) do
18.     if (!BackScan(S_p^i, SDB^S_pi))
19.    BEI =backward extension check(S_p^i, SDB^S_pi);
20.    call bide(SDB^S_pi, S_p^i, min_sup, BEI, FCS);
```

We next summarize how BIDE works. Then, we present our extensions on it in order to mine positional sequence patterns. BIDE focuses on a special class of patterns, called 'closed' patterns. Algorithm 1 summarizes the working principles of BIDE. We first provide a definition of 'closed' pattern.

**Def'n** *(Regular Closed Sequence Pattern):* Assume that $S_a$ and $S_b$ are two regular sequence patterns. If $S_b$ properly contains $S_a$, then $S_b$ is called a *supersequence* of $S_a$. If a sequence pattern has no *supersequence* with the same frequency, then it is called a *regular closed sequence pattern*.

**Example**: Consider the following sequence patterns and their frequencies: $S_a$ = ABC (freq: 5), $S_b$ = ABBC (freq: 5), $S_c$ = ABB (freq: 6). Here, $S_a$ is not a regular closed pattern, since $S_b$ is a supersequence of $S_a$, and it has the same frequency. On the other hand, although $S_b$ is also a supersequence of $S_c$, $S_c$ is still a regular closed pattern, as its frequency is higher than $S_b$.

**Def'n** *(First instance of a prefix sequence):* Given an input sequence *t* and a prefix sequence *s*, if *t* contains *s*, the subsequence from the beginning of *t* to the end of the first appearance of *s* is called the *first instance of prefix sequence* in *t*. As an example, the first instance of prefix sequence *CD* in sequence *ABCDAB* is *ABCD*.

**Def'n** *(Projected sequence of a prefix sequence):* The projected sequence of a prefix sequence is the remaining part of the input sequence *t*, after removing the first instance of the prefix sequence *p* from *t*. As an example, the projected sequence of prefix sequence *CD* in sequence *ABCDAD* is *AD*.

**Def'n** *(Projected database of a prefix sequence):* The projected database of a prefix sequence *t* in a database *D* is the complete set of the projected sequences of *t* in *D*.

Given a sequence database, BIDE first scans the entire database to find all frequent sequences that have length 1 in line 1. It then builds the projected database for each frequent length-1 sequence from line 3 to 4. Next, it call BackScan to check whether frequent 1-sequence can be pruned or not in line 6 and if not it computes the number of backward-extensions items in line 7 and call the subroutine bide in line 8. In line 10, it finds all locally frequent items for a prefix and compute the number of forward-extension items in line 11, and if no forward-extension and backward extension, it sign prefix as frequent closed sequence in line 12 to 13 and extend prefix to get new prefix in and find pseudo projected database for new prefix in line 15 and 16 and then it applies BackScan check from line 17 to 19 and apply subroutine bide in line 20. Figure 2 summarizes the workflow of BIDE for the frequent length-1 sequences (the produced patterns are presented in SQL LIKE syntax).

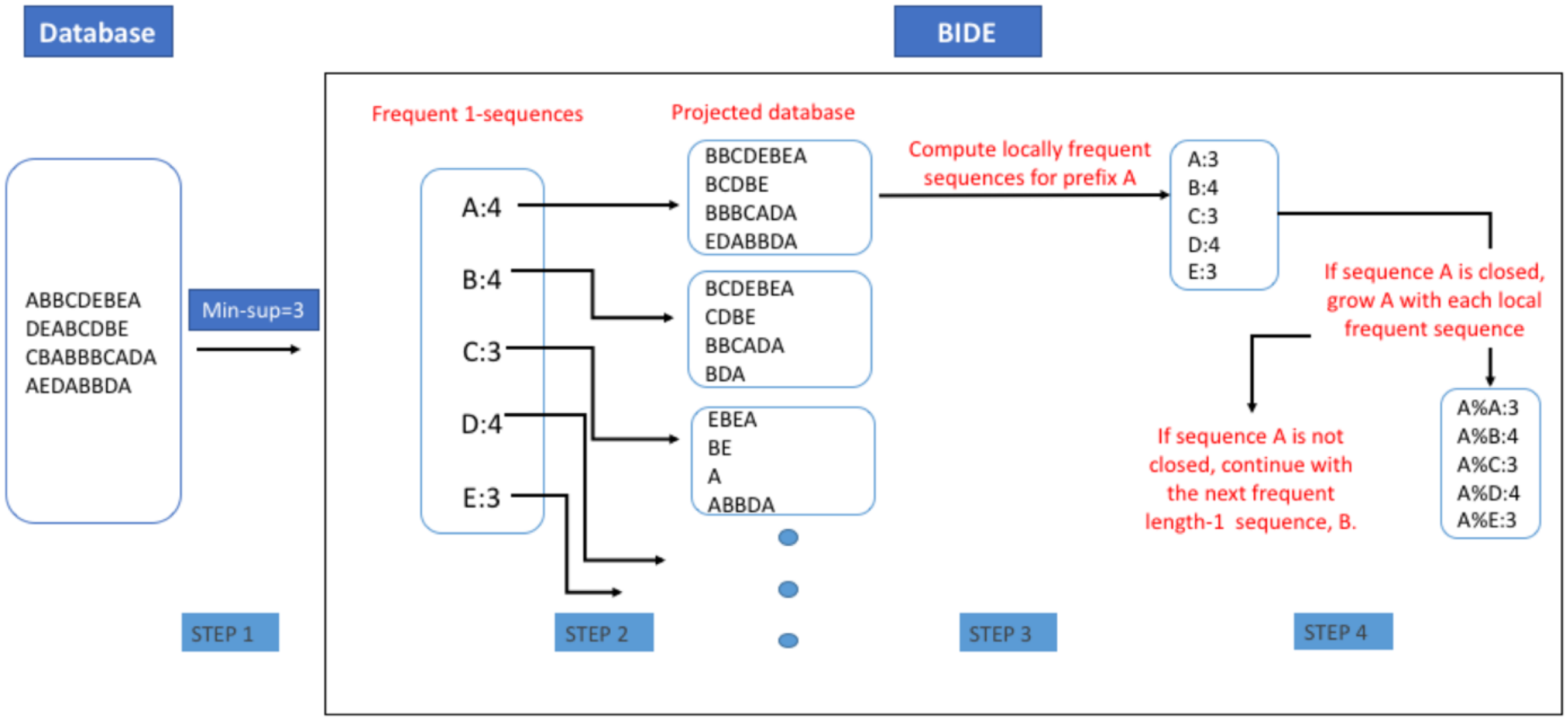

Figure 2: A high level overview of BIDE steps for length-1 patterns

### *3.1.1 Extensions for Positional Sequence Pattern Computation*

In order to mine positional sequence patterns, we extend step 3 in Figure 2 and line 10 in algorithm 1. In step 3, all locally frequent items are computed for a prefix sequence in the corresponding projected database. That is, BIDE goes over all sequences in a projected database, and counts whether an item of interest exists in the projected sequence or not. Our extension divides this step into two parts. In the first part, locally frequent items are computed in the same way as in BIDE. In the second part, we compute the locally frequent sequences where there is no item between the last item of the prefix sequence and the local item. Figure 3 illustrates the new workflow in step 3 of the extended algorithm for the database illustrated in figure 1 (the produced patterns are presented in SQL LIKE syntax).

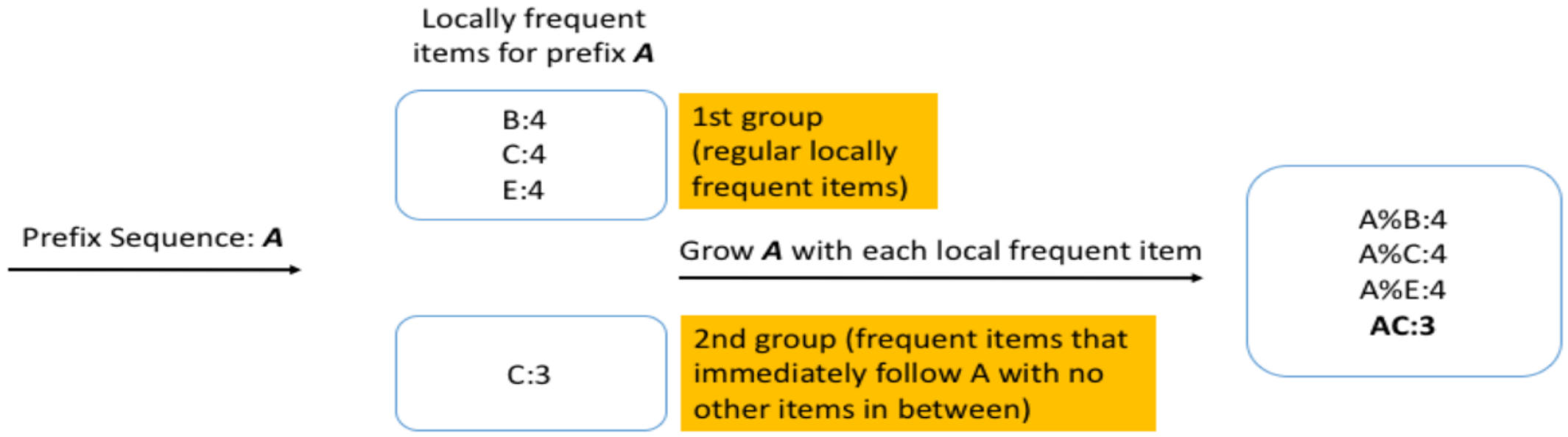

Figure 3: The new workflow in step 3 of the extended algorithm

In Figure 3, there are two groups of locally frequent items for prefix *A*. The first group includes the all regular locally frequent items as before. The second group includes only locally frequent items which do not have any characters between the last character of the prefix (frequent closed sequence) and the first character of the projected sequences for that prefix. When prefix A is extended with these locally frequent items, the new prefix sequences are {*A%B*:4, *A%C*:4, *A%E*:4, ***AC*:3**} where the last sequence is obtained from the second group of locally frequent items.

Note that a locally frequent item may appear in both groups 1 and 2. In such cases, after verifying that their frequencies are over the minimum support threshold, we grow the prefix sequence as follows:

- If both frequencies are equal, extend the prefix sequence with only the local item from the second group (closure check).
- Otherwise, extend the prefix sequence with both local items.

The above first bullet-point enforces the "closed pattern" property. That is, since the patterns extended with locally frequent items from the second group are more specific than their counterparts in group 1, the former ones are preferred over the latter to make sure that the computed patterns are closed.

### 3.2 Histogram Construction

Out of the mined positional sequence patterns, a histogram is built in a similar way as in SPH (Aytimur & Çakmak, 2018) with some extensions. We first briefly explain the histogram building steps (please refer to (Aytimur & Çakmak, 2018) for details), and then present our extensions to eliminate redundant endpoints.

In order to build a histogram, first, all sequential patterns are lexicographically sorted. The capacity of each bucket is determined by dividing the total frequencies of the patterns by the number of buckets. In the next step, bucket endpoints are determined. To this end, the sorted pattern list is traversed starting from the first one while keeping a running sum S of pattern frequencies. Whenever $S \geq n * C$, where n is the currently built bucket's number with initial value 1, and C is the bucket capacity, that particular pattern is set as the current bucket's endpoint, and n is updated as $n = floor(S/C) + 1$. Then, this process is repeated until all the histogram endpoints are determined. We give an example below.

**Example:** Consider the set of positional sequence patterns that are mined with minimum support threshold 3 from the database in Figure 1, i.e., {**A%C%BE**:3, **AC%B**:3, **AC%E**:3, A%C%B:4, A%C%E:4, B%A%B:3, B%A%E:3, B%C%B:3, B%C%E:3, B%E:4, C%C%B:3, C%C%E:3}. Assume that the specified bucket count is 5. Since the total pattern frequency is 39, the capacity for each bucket is 7.8. Figure 4 shows the histogram constructed by employing the above summarized approach.

| Endpoint Number | Endpoint Value | Endpoint Frequency |
|---|---|---|
| 12 | **AC%E** | 3 |
| 22 | **A%C%E** | 4 |
| 32 | B%C%B | 3 |
| 42 | B%E | 4 |
| 48 | C%C%E | 3 |

Figure 4: Positional pattern-based histogram construction with bucket count 5

#### *3.2.1 Eliminating Redundant Endpoints*

We observe that many endpoint patterns in the constructed histograms are highly similar with small difference in their frequencies. Such patterns occupy space in scarcely available histogram endpoints, while not providing much extra coverage in terms of matching a diverse set of LIKE query predicates. We give an example.

**Example**: Consider two positional frequent sequences, A%C%BE and AC%B, with the same frequency. Almost all query predicate patterns that match to the second pattern also match to the first pattern. Therefore, it would be redundant to keep both patterns in the histogram. Instead, the first pattern may be kept, as it differentiates between a wider spectrum of possible query predicate patterns, and the second pattern may be discarded from the histogram.

As an extension to SPH (Aytimur & Çakmak, 2018), we eliminate such patterns based on pattern containment and information theoretic filtering as formally defined next.

**Def'n (Pattern Containment)**: Given a positional sequence pattern P, let *Striped(P)* denote a sequence which contains all non-wildcard characters of P in the same relative order as in P. Then, a positional sequence pattern Q is *contained* in another positional sequence pattern S, if *Striped(Q)* is *properly contained* in *Striped(S)*.

In order to eliminate the adverse effects of the redundant pattern issue, one may consider eliminating a pattern if it is contained in at least one other pattern. However, such an approach may lead to major estimation errors, as the "information" contained in these patterns may differ significantly. Hence, we propose to compute the *information content* (Cover & Thomas, 2012) of patterns, and eliminate a pattern *p*, if the information content difference between *p* and another pattern that *contains p* is "ignorably" small.

**Def'n (Information Content of a Pattern)**: Given a positional sequence pattern R and a database D, let *freq(R)* denote the number of rows that contain R in D, and |D| denote the number of rows in D. Then, the information content of R, IC(R), is computed as follows:

$$IC(R) = -\,log\;log\;P(R)$$

where $P(R)$ denotes the probability of R, and computed as $P(R) = \frac{freq(R)}{|D|}$.

**Def'n (Redundant Pattern)**: Given a positional sequence pattern set S computed over a database, a pattern P ∈ S is considered *redundant*, if there exists another pattern R ∈ S and R ≠ S such that (i) R *contains* S, and (ii) $IC(R) - IC(S) < \delta$ where $\delta$ is a small threshold.

In the above definition, the information content difference threshold, $\delta$, is determined experimentally as explained in the empirical evaluation section. Before histogram computation, we eliminate *redundant patterns*. Then, the remaining patterns are considered during histogram construction. Once a pattern-based histogram is built offline during database statistics gathering time, it is stored in the database dictionary/catalog to be later used during query optimization time for selectivity estimation.

## 4. LIKE Predicate Selectivity Estimation with Pattern-based Histograms

We estimate the selectivity of LIKE query predicates based on the constructed histogram in a similar way to SPH (Ayimur & Çakmak, 2018) with some extensions. At a high level, the selectivity of a LIKE predicate pattern p is estimated according the type of the match between predicate p and histogram endpoints as summarized below. Note that the order is important, i.e., exact match is preferred over encapsulated match, if both are applicable.

- p *exactly matches* a histogram endpoint b. [*Exact match case*]
  - $selectivity = the\ endpoint\ frequency\ of\ b\ /\ the\ database\ size$
- p is *contained* in a set B of histogram bucket endpoints. [*Encapsulated match case*]
  - $selectivity = the\ minimum\ endpoint\ frequency\ in\ B\ /\ the\ database\ size$

where set B includes bucket endpoint values and their corresponding frequencies for buckets in histogram that encapsulate p and minimum endpoint frequency is the minimum endpoint frequency value in that set B.

- otherwise: [*No match case*]
  - $selectivity = t\%\ of\ the\ minimum\ support\ threshold/\ the\ database\ size$

where SPH determines t experimentally as 10%, and we use the same setting.

In the above approach, due to the limited number of histogram endpoints, many queries fall into "no match case" in which an average selectivity is assigned independent of the query predicate. On the other hand, it is often the case that some parts of the query predicate may match to the histogram endpoints, which may provide better estimation boundaries. Hence, in addition to the above matching approaches of SPH, we introduce a *partitioning-based matching* strategy that is employed when exact or encapsulated matching attempts fail as summarized next.

### 4.1 Partitioning-based Matching

The idea behind partition-based matching is stated by the following Lemma.

**Lemma (Substring Selectivity)**: Given a LIKE query predicate string S of length n, assume that S is divided into two pieces, $S_{1..i}$ and $S_{i+1...n}$, at position i. Then, $Selectivity(S) \leq Min[Selectivity(S_{1..i}), Selectivity(S_{i+1..n})]$.

The proof of the above lemma is straightforward and intuitive. Therefore, we omit a formal proof here for brevity. We perform the partition-based matching by introducing a slider, initially positioned at the first character in a given query predicate string S. Then, we advance the slider position one-by-one until it reaches to the last character in S (Figure 5). At each slider position, we attempt to compute the selectivity estimates of $S_{1..i}$ and $S_{i+1...n}$ with exact or encapsulated match options. Next, the computed partition selectivities are compared to the minimum of the previously computed selectivities of partitions for the earlier positions of the slider. If any of the selectivities computed for the current position of the slider is smaller than the currently known minimum selectivity among all previous partitions, then we update the minimum selectivity accordingly. Once the slider reaches position n, the minimum selectivity estimate computed over all slider positions is assigned as the selectivity estimate for S.

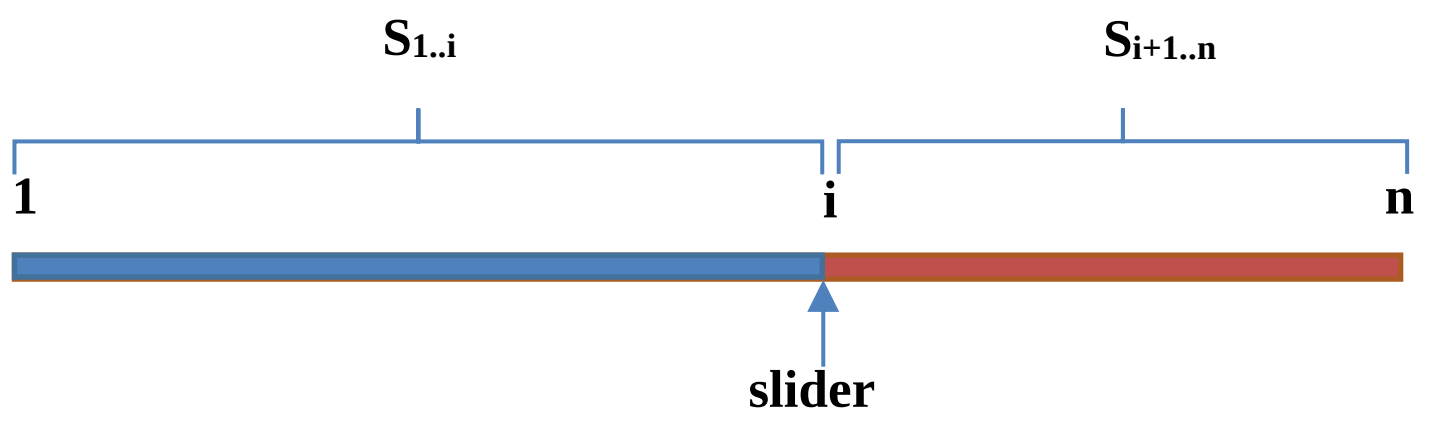

Figure 5: Slider-based partitioning of query predicate string S

***Partition length-based filtering*:** We observe that bluntly employing the above approach may sometimes lead to highly overestimated selectivity predictions. This happens in the following cases: a very short partition (e.g., one character) of a query string may have exact or encapsulated match in the histogram with many high frequency endpoints, and the remaining partitions may not have any match. Hence, the selectivity of that very short partition is used as the selectivity estimate (as it is the "minimum" among all partitions). However, since there is a large length difference between the original query predicate string and its short partition(s), their selectivities usually differ quite a bit as well. In order to alleviate this side-effect of partitioning, we eliminate partitions of length smaller than $n - \varepsilon$ from consideration, where n is the length of the query predicate string, and $\varepsilon$ is an experimentally determined threshold value. In the experimental evaluation section, we study the effect of $\varepsilon$, and accordingly, determine its optimal value.

The partitioning-based matching is placed after the *encapsulated match case* in the above list of attempted matches in the original SPH (Aytimur & Çakmak, 2018) approach. That is, if the exact and encapsulated match is not possible, then the partitioning-based match is explored. If slider-based partitioning does not yield any selectivity estimate either (i.e., none of the partitions qualify for exact or encapsulated match), then 'no match case' is employed as before. We next give an example that comprehensively illustrates the effect of new histograms and partition-based matching on the selectivity estimation process.

**Example:** Consider the database D in Figure 6 that contains 8 rows. Assume that the minimum support threshold is 2, and the allowed number of buckets in a histogram is 4. 12 regular sequence patterns and 18 positional sequence patterns are computed from D. Figure 7 shows the resulting histogram constructed from regular sequence patterns, and Figure 8 shows the histogram constructed from positional sequence patterns. Note that the histogram built from positional sequence patterns (Figure 8) is more specific than the one constructed from regular patterns (Figure 7). As an example, the endpoint pattern for bucket 2 in Figure 7 has two forms in the other histogram in Figure 8. We next demonstrate the selectivity computation for the above-discussed four match cases with both histograms.

| Sequence id | Sequence |
|---|---|
| 1 | BCAB |
| 2 | BCAC |
| 3 | ACXCB |
| 4 | ACYCB |
| 5 | ACZCB |
| 6 | AYCCB |
| 7 | AZCCB |
| 8 | ACXCB |

Figure 6: An example sequence database D

| Endpoint Number | Endpoint Value | Endpoint Frequency |
|---|---|---|
| 15 | A%B | 7 |
| 28 | A%C%C%B | 6 |
| 42 | B | 8 |
| 59 | C%B | 7 |

Figure 7: Histogram constructed from regular sequence patterns

| Endpoint Number | Endpoint Value | Endpoint Frequency |
|---|---|---|
| 20 | A%C%CB | 6 |
| 36 | AC%CB | 3 |
| 54 | C | 8 |
| 74 | C%CB | 6 |

Figure 8: Histogram constructed from positional sequence patterns

*Exact match case:* Assume that a LIKE query predicate p = A*C%CB* is given. SPH (Aytimur & Çakmak, 2018) would declare *exact match* to the second bucket of the histogram in Figure 7, and compute the selectivity as 0.75 (i.e., 6/8 where 6 is the corresponding bucket endpoint frequency, and 8 is the number of rows in the database). Similarly, P-SPH would declare *exact match* to the second bucket of the histogram in Figure 8, and compute the selectivity as 0.375 (i.e., 3/8 where 3 is the corresponding bucket endpoint frequency, and 8 is the number of rows in the database). The true selectivity for p is 0.375.

*Encapsulated match case:* Assume that a LIKE query predicate p = *%C%C%* is given. No *exact match* is possible in either of the histograms. However, there is an *encapsulated match* with the second bucket of the histogram in Figure 7. Since it only matches with a single bucket, SPH's selectivity estimate is 0.75 (i.e., 6/8). As for the histogram in Figure 8, there are encapsulated matches with buckets 1, 3, and 4, where the maximum endpoint frequency is 6. Thus, the selectivity estimate of P-SPH is 0.75 (i.e., 6/8) as well. The true selectivity for p is 0.875 (i.e., 7/8).

*Partitioned Encapsulated match case:* Assume that a LIKE query predicate p = *Z%CB* is given. No exact or encapsulated match or encapsulated match is possible in any of the histograms. SPH will employ the no match case, and estimate the selectivity as 0.025 (i.e., 0.2/8) (0.2 is *t% of min support threshold* where SPH uses an experimentally determined value 10 for t). P-SPH takes advantage of partitioning-based match in this case. According to Figure 8, partitioned match is possible between p and the first, second and fourth endpoint values. Out of these, since the second bucket endpoint provides the smallest estimate, the selectivity is 0.375 (3/8). The true selectivity for p is 0.25 (2/8).

*No match case:* Assume that a LIKE query predicate p = *D%A%D%E* is given. In this case, none of the above match types (including partition-based match) is possible in either histogram. Therefore, *no match case* route is taken, and the selectivity is estimated as 0.025 (i.e., 0.2/8). The true selectivity for p is 0.

## 6. Experimental Results

This section presents our experimental results to evaluate the proposed P-SPH algorithm. All of our experiments were performed on a DELL R720 machine with 2 x XEON E-5-2620v2 2.10 GHz CPU and 80 GB of RAM.

### 6.1 Dataset

We perform various experiments using a real dataset from DBLP. The dataset is the same as the one used in SPH (Aytimur & Çakmak, 2018) and contains 800,000 full author names. The lengths of full author names vary between 18 and 60 with an average of 22.5.

### 6.2 Test Query Set

We evaluated the accuracy of P-SPH using the same query workload described in SPH (Aytimur & Çakmak, 2018). More specifically, the query workload includes three different groups of queries, and each group has 100 queries, except that the negative query set contains 24 queries. The way that these query sets are generated are described in SPH (Aytimur & Çakmak, 2018), but we also include a brief summary here as well.

- Queries in the first group are in the form of *%w% and* $\%w_1\%w_2\%$ where $w_i$ is a word that has length between 5 and 12. In order to generate this group of queries, one or two words with length between 5 and 12 are chosen. Then, a random number (from 0 to 2) of underscore characters (i.e., "_") are inserted at random positions in a word. In SQL LIKE syntax, the underscore character represents a wildcard that matches any single character. This group of query predicates has minimum, average, and maximum length of 5, 6.7, and 17, respectively. The average selectivity is 4.77%.
- To construct the second group of the queries, a random row, R, from the database is chosen, and a random number, k, between 3 and the length of this selected row is drawn. Then, k characters are removed from R, and a random number (from 2 to 8) of "%" signs are inserted at random positions in R. Clearly, the generated queries are in the form of $\%s_1\%s_2\%.... \%s_n\%$ where $s_i$ represents one or more characters. The average, minimum, and maximum lengths for the query predicates in this set are 8.4, 3, and 16, respectively. The average selectivity is 2.67%.
- Negative query set includes queries which do not match any rows in the database. The generation of the negative queries is almost the same as second group of queries. The only difference is that a smaller number (from 1 to 3) of "%" symbols are randomly inserted in R to decrease the likelihood that the generated query predicate match to any rows in the database. Out of 100 generated queries, 24 of them are truly negative queries with 0 matching rows. Hence, this set, in its final form, contains 24 queries.

### 6.3 Evaluation metrics

SPH (Aytimur & Çakmak, 2018) employs two metrics to test the accuracy of selectivity estimation. The first metric is the relative error which is employed for the positive query sets (the first and second query groups above). The relative error is defined as $|f_{true} - f_{est}| / f_{true}$, where $f_{true}$ is the actual true selectivity of the query, and $f_{est}$ is the estimated selectivity. Since the third group queries (negative query set) has the actual selectivity of 0, the relative error metric is not applicable here (i.e., due to the division by 0 error). Instead, the *absolute error* metric is used in this group. Absolute error is defined as $|f_{est} - f_{true}|$. We use the same metrics in order to evaluate our technique and compare it with the state of the art SPH (Aytimur & Çakmak, 2018). Moreover, in SPH, the authors exclude those queries that have actual frequency of 10 or less. Similarly, we also exclude such queries in this work as well.

### 6.4 Results

In this section, we evaluate different aspects of our approach in terms of estimation accuracy, query time, space overhead, and compare it to the state of the art.

#### *6.4.1 The effect of the minsup threshold*

Minimum support threshold, during frequent sequence mining, directly affects the number of patterns in the result set. This section evaluates the effect of the minimum support threshold on the computed pattern count and accuracy. Figure 9 shows the number of the regular and positional sequence patterns for different minimum support values.

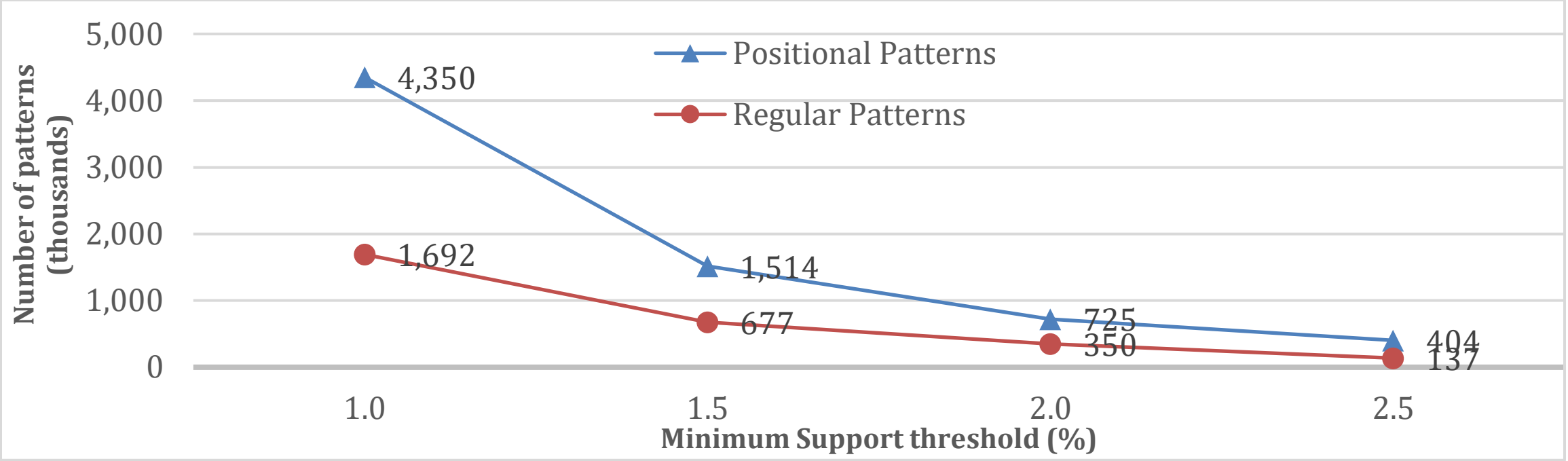

Figure 9: The number of regular and positional sequence patterns for different minsup values

***Observation 1****: The total number of patterns (positional or regular) decreases dramatically, as the minimum support threshold increases.*

The total number and the rate of increase for positional patterns is slightly higher than the regular patterns. The reason for this difference is that positional sequence pattern set includes all regular sequence patterns as well as some additional patterns. For instance, consider two patterns and their frequencies: A%B%A%D%A%B: 5 and A%B%A%D%AB:3. Regular sequence pattern set includes only the first one, while the positional sequence pattern set includes both.

As per the above observation, since the minimum support threshold greatly affects the total number of patterns, it is critical to increase the minimum support threshold to the extent that it does not harm the selectivity estimation accuracy considerably. Next, we investigate how high the minimum support threshold could be while keeping the selectivity estimation accuracy decrease tolerable. Figure 10 shows the change of accuracy with different minimum support threshold values (bucket count: 2048).

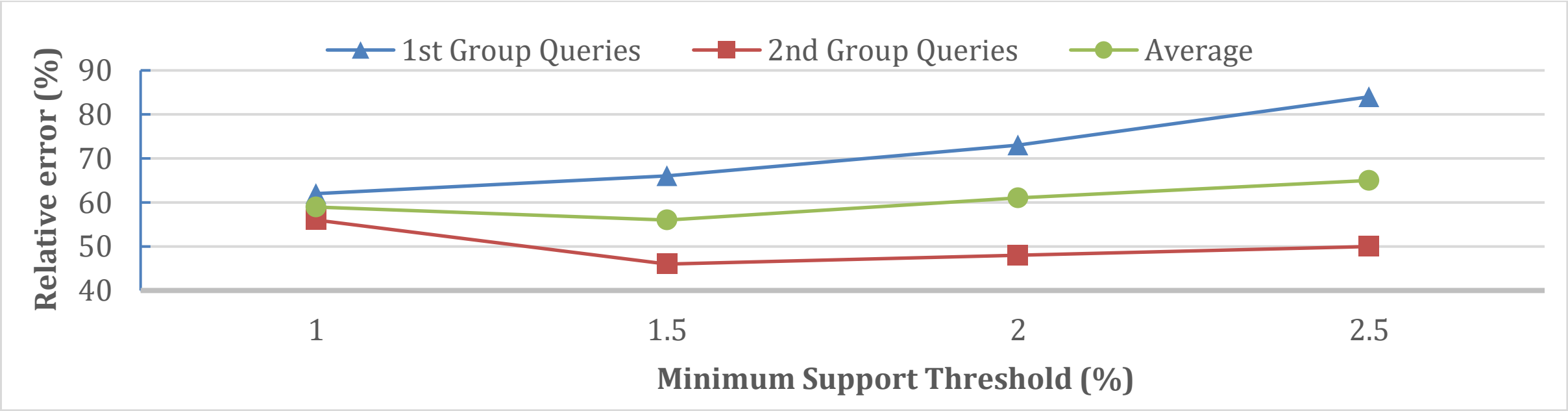

Figure 10: The change of accuracy with different minsup values

***Observation 2****: The average accuracy gets the highest value when the minimum threshold equals to 1.5%.*

The accuracy of estimation decreases in general, as the minimum support threshold increases. The reason is that as the minsup threshold gets larger, the mined patterns get shorter. Hence, the ratio of queries that can be answered with exact or encapsulated match decrease, and P-SPH more often uses partitioning-based and no-match cases, which are more error prone. There is one exception to this expected behavior. That is, the second group queries experience decrease in error rate when minsup threshold increases from 1% to 1.5%. The reason is that for the first group of queries, the number of non-matching queries increase as the minsup value increases; hence, the error rate increases as well. On the other hand, for the second group of queries, the number of non-matching queries stays the same when minsup changes from 1% to 1.5%. What is more, the average actual row count for non-matching queries decrease, which closes the distance between the fixed estimate (minsup*10%) that is used in *no match* case. Therefore, the error rate decreases for 2nd group queries when minsup increases from 1% to 1.5%. Detailed values for each minsup value are presented in Table 1.

| *Minsup* | 1% | | 1.5% | | 2% | |
|---|---|---|---|---|---|---|
| *Query Group* | 1st group | 2nd group | 1st group | 2nd group | 1st group | 2nd group |
| *Number of non-matching queries* | 42 | 25 | 49 | 25 | 52 | 30 |
| *Average row count for non-matching queries* | 4288 | 9978 | 5657 | 8448 | 5645 | 10801 |

Table 1: Non-matching query statistics for each minsup value

Based on the above observation, unless noted otherwise, we use 1.5% as the minimum support threshold for the remaining experiments.

### *6.4.2 The effect of the number of buckets*

The number of buckets in a histogram is an important setting, as more number of buckets means more endpoint values, and may increase the selectivity estimation accuracy. However, lower numbers of buckets require less search time and memory. Hence, there is a trade-off between the estimation accuracy and higher resource consumption. In this experiment, we investigate the relationship between the selectivity estimation accuracy and the number of buckets. Figure 11 plots the relative estimation error for different numbers of buckets.

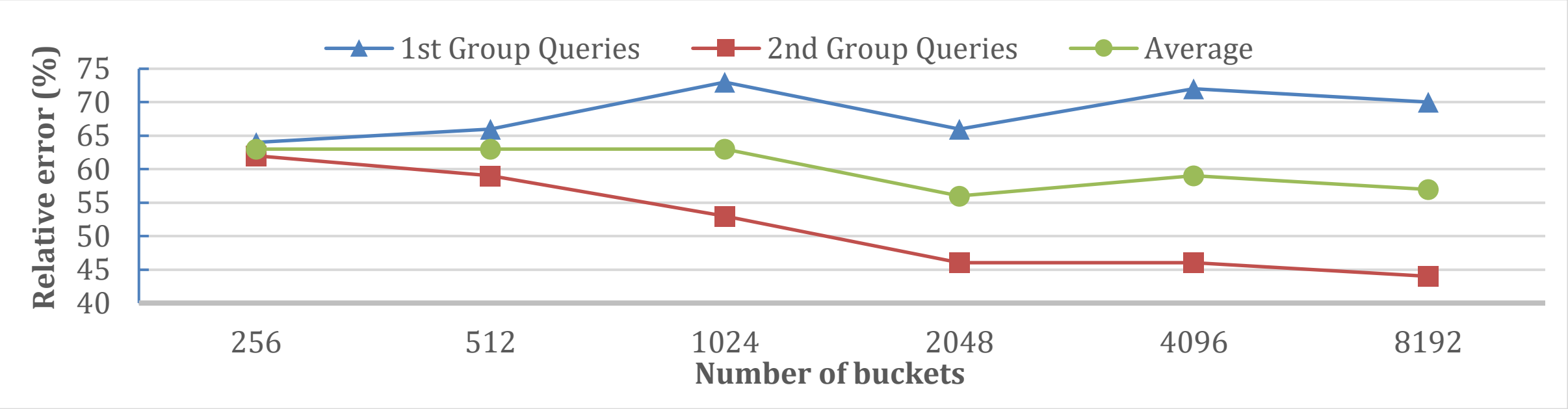

Figure 11: Relative error for different numbers of buckets for P-SPH with minsup 1.5%

***Observation 3****: As the number of buckets increase, the average relative error decreases until the number of buckets equal to the 2048. After that point, no meaningful accuracy improvement is observed.*

The above observation shows that there is no need to use more than 2048 buckets to increase the selectivity estimation accuracy. Hence, unless noted otherwise, we use 2048 buckets in the remaining experiments.

### *6.4.3 The effect of partitioning*

In this section, we evaluate the contribution of partitioning-based matching. First, we determine the optimal value for the threshold $\varepsilon$ that provides the best accuracy. Figure 12 shows the change of accuracy for different values of $\varepsilon$ (bucket count: 2048, minsup: 1.5%).

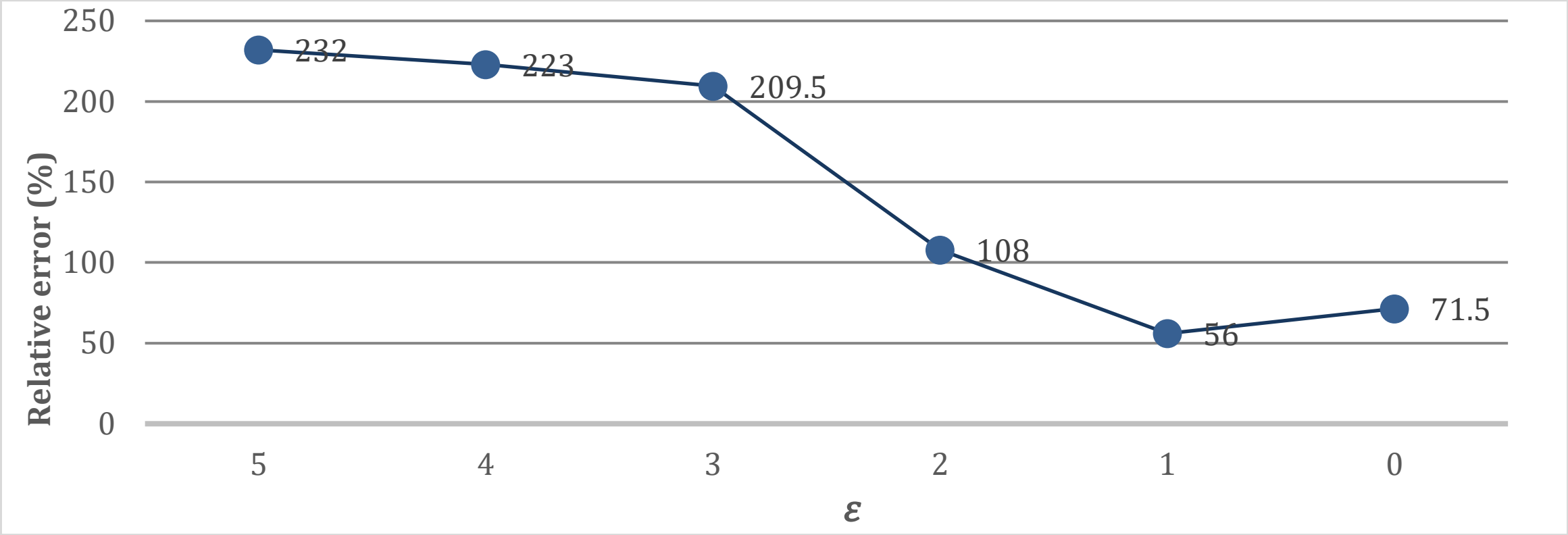

Figure 12**:** Average relative error for different values of $\varepsilon$

***Observation 4****: The lowest relative error is obtained when $\varepsilon$ is 1, i.e., partitions which are shorter than $|length\ of\ query\ predicate| - 1$ are not considered during partitioning-based selectivity estimation.*

Relative error increases when the allowed partition lengths get shorter. This is because short partitions are very generic, and their selectivity may be a lot higher than the original predicate. Hence, the relative error increases.

Next, we present selectivity estimation accuracies with and without partitioning-based matching is used. Figure 13 shows the average relative error for different minimum support threshold values with and without partitioning (bucket count: 2048, $\varepsilon = 1$).

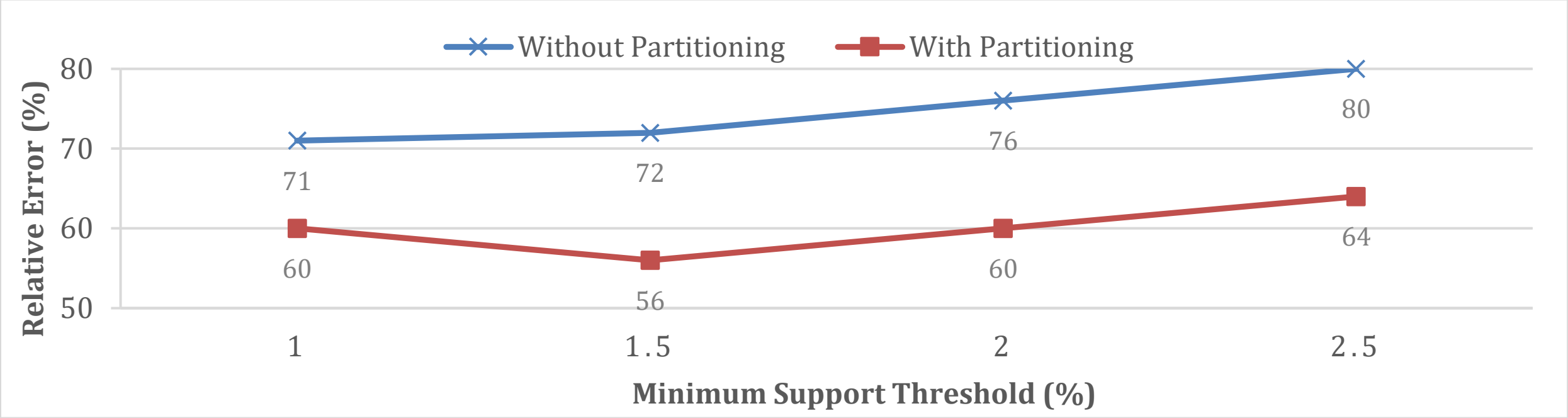

Figure 13**:** Relative error for different minimum support thresholds values with and without partitioning

***Observation 5****: Partitioning-based matching decreases the relative selectivity estimation error up to 16%.*

When partitioning-based match is not available, long query predicate strings do not match any endpoint of the histogram, although there are matching rows in the database. On the other hand, partitioning query predicate strings allow such queries to match histogram endpoint values, and provides better estimation accuracy.

Finally, we evaluate the impact of partitioning-based matching on query processing time. Figure 14 shows the average selectivity estimation time with and without partitioning (bucket count: 2048, minsup: 1.5%).

***Observation 6****: Partitioning-based matching increases the average running time around 3 times.*

The increase in query processing time due to partitioning-based matching is expected, as it adds an extra step of computation during the selectivity estimation. Although the selectivity estimation time relatively increases significantly, the total estimation time is still at the level of milliseconds. Thus, the improved accuracy as shown in Figure 12 justifies this estimation time compromise.

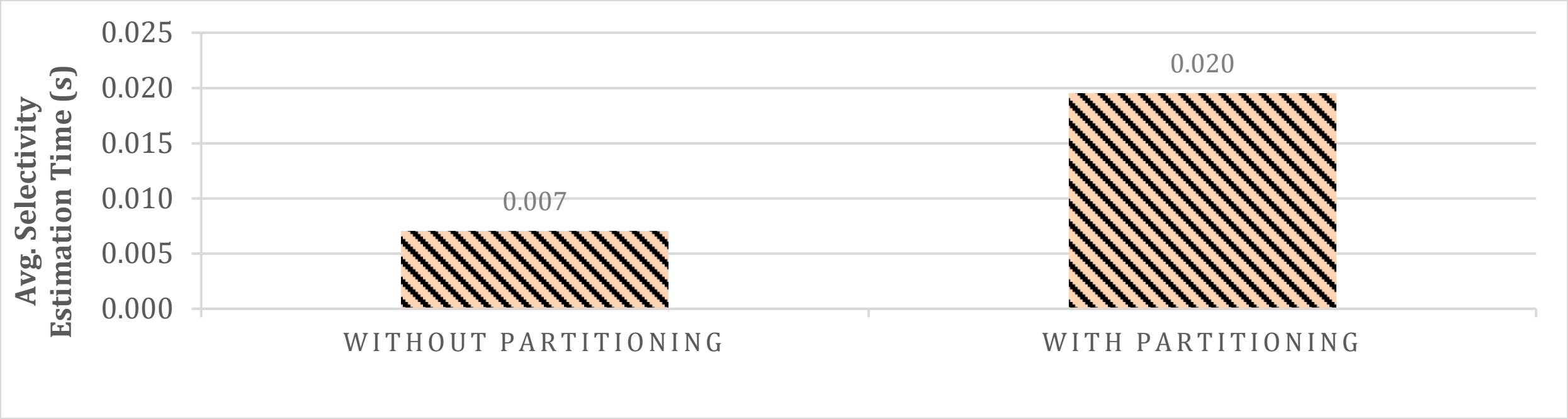

Figure 14: Selectivity estimation time with and without partitioning

### *6.4.4 The effect of redundant pattern elimination*

In this section, we evaluate the effect of redundant pattern elimination. We first experimentally identify the best value for the information content difference threshold, $\delta$, described in section 3.3.1. Figure 15 shows the change of relative estimation error with different information content difference threshold values where the horizontal axis values are multiplied by $10^5$ to improve the readability (bucket count: 2048, minsup: 1.5%).

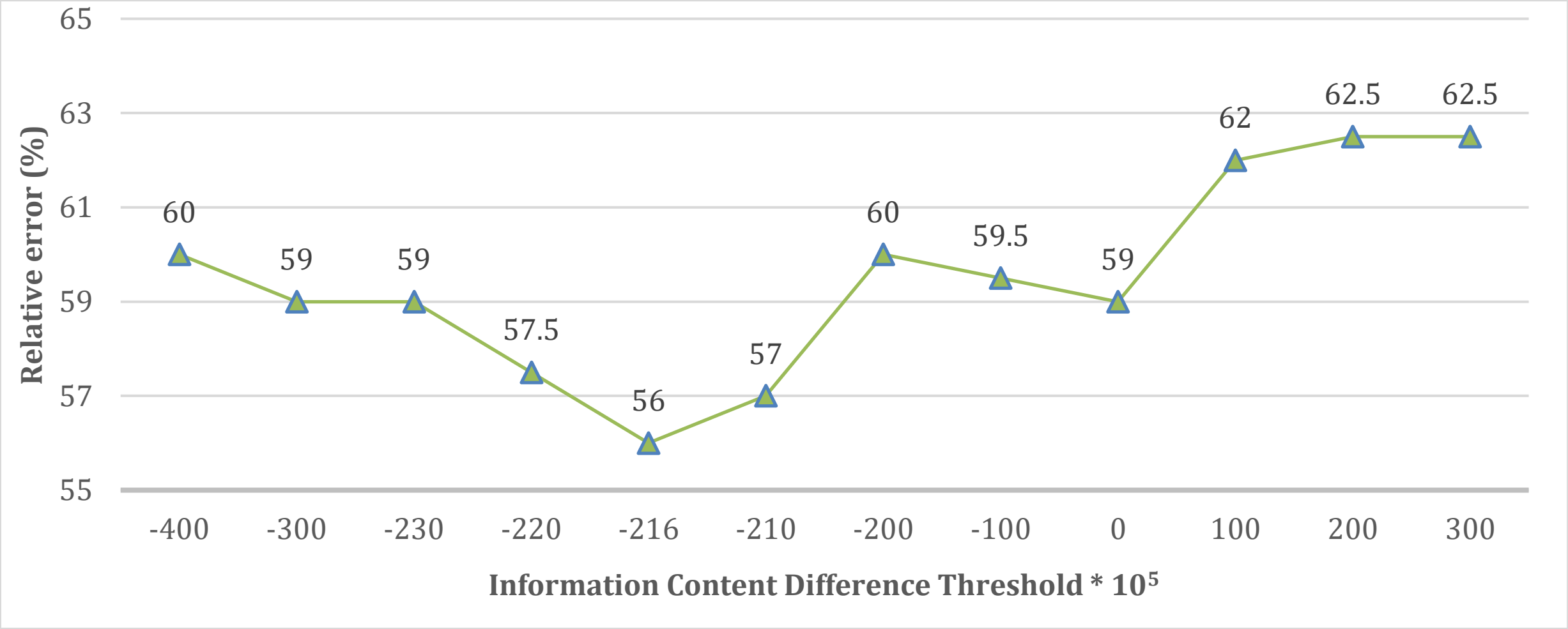

Figure 15: The change of relative estimation error with different information content difference threshold values

***Observation 7****: The lowest relative error is obtained when the information content difference threshold for redundant pattern elimination is -0.00216.*

Based on the above observation, for all the experiments in this paper, we set the information content difference threshold for redundant pattern elimination as -0.00216.

Next, we present the impact of redundant pattern elimination on the selectivity estimation accuracy. Figure 16 shows the average relative error for different minimum support threshold values with and without redundant pattern elimination (bucket count: 2048, $\varepsilon = 1$).

***Observation 8****: Redundant pattern elimination decreases the relative error up-to 7%.*

Redundant pattern elimination does not harm the estimation accuracy, i.e., it either improves the accuracy or performs nearly the same as the case without redundant pattern elimination. This is expected, as redundant patterns occupy some of the very limited histogram bucket endpoints, and removing them opens up space for more distinctive patterns to be included in a histogram.

Finally, we evaluate the impact of redundant pattern elimination on histogram construction time. Figure 17 shows the change of histogram construction time with and without redundant pattern elimination (bucket count: 2048).

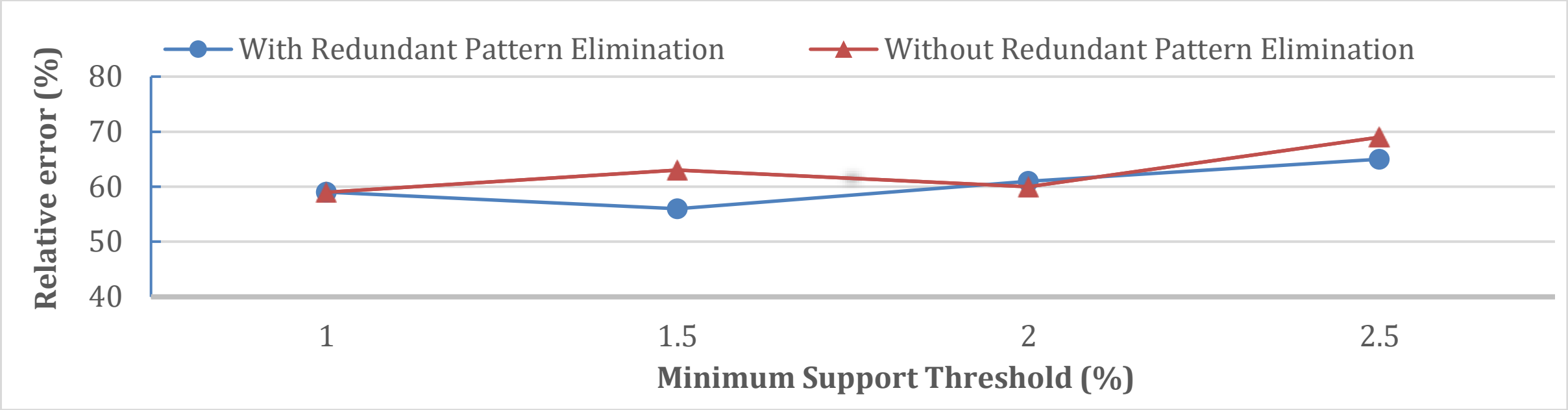

Figure 16**:** Average relative error with and without redundant pattern elimination

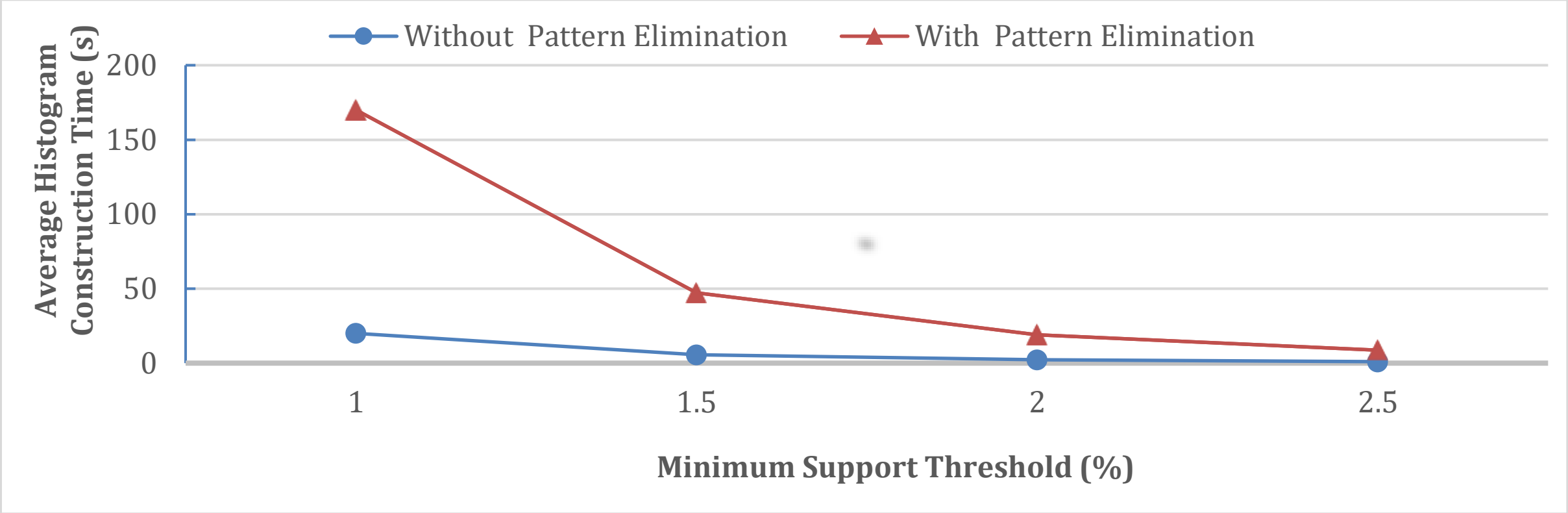

Figure 17: Build time with and without redundant pattern elimination

***Observation 9****: Redundant pattern elimination significantly increases the histogram construction time.*

Although the above observation points to some considerable overhead in histogram construction time, the bottleneck in the statistics gathering pipeline is the pattern mining stage, which is responsible for more than 99% of the total spent time as shown in the next section. Therefore, the above increase in histogram construction time is invisible in practice.

***6.4.5 Comparison with the state of the art***

In this section, we compare our proposed method (P-SPH) with the state of art, SPH [6] in different aspects, such as accuracy, time and space requirements, etc. Also, we compare accuracy of P-SPH with two different approaches in the literature, KVI [2] and LBS [1] algorithms.

*Accuracy-based comparison:*

We first conduct a comparison study based on the selectivity estimation accuracy. SPH originally does not have partition-based matching (PBM) and redundant pattern elimination (RPE) features. On the other hand, these techniques are generic enough to be transferred from P-SPH to SPH. Hence, for comparison purposes, we also create an enhanced version of SPH extended with PBM and RPE. Figure 18 shows the average accuracy values for the original SPH, extended SPH, and P-SPH approaches with different minsup thresholds (number of buckets: 2048).

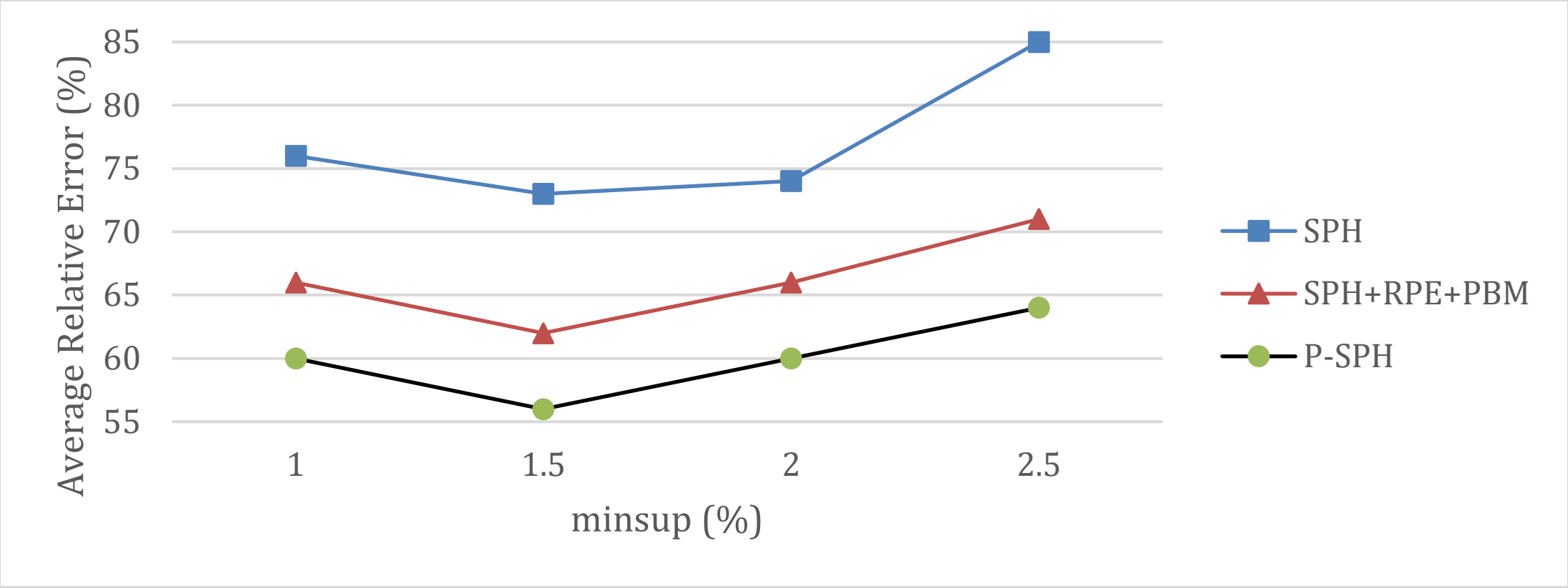

Figure 18: Average relative error comparison for different minimum support thresholds values

***Observation 10**: P-SPH provides up to 20% lower error rates in comparison to the original SPH.*

***Observation 11**: P-SPH provides up-to 7% better selectivity estimation accuracies than the extended SPH for all minsup values. Since the only difference between P-SPH and the extended SPH is that the former employs positional sequence patterns while the latter employs regular sequence patterns, the success of P-SPH over the extended SPH is attributed to the positional sequence pattern usage.*

We next perform a similar comparison by changing histogram bucket counts and keeping the minsup value fixed this time. Figure 19 shows the average accuracy values for the above three approaches with different bucket counts (minsup: 1.5%).

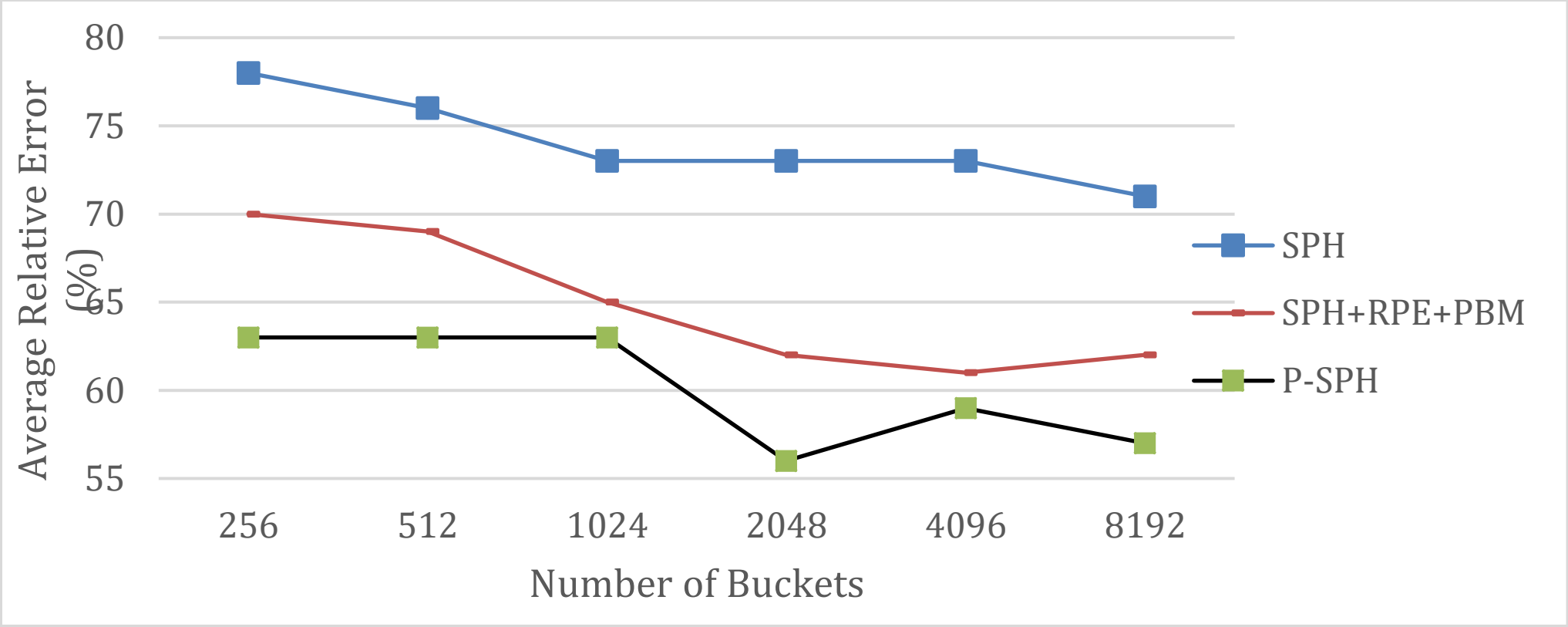

Figure 19: Relative error comparison for different minimum support thresholds values

***Observation 12**: In all bucket count configurations, P-SPH outperforms SPH (up-to 17%). Both approaches, in general, benefit from increased bucket counts in terms of lowering the estimation error rates.*

***Observation 13**: Positional sequence patterns lead to better selectivity estimates than the regular sequence patterns as P-SPH outperforms (up-to 7%) the extended SPH featuring partitioning-based matching and redundant pattern elimination for all bucket configurations.*

The above two experiments show that P-SPH outperforms both SPH and its extended version for all configurations of minsup and bucket counts.

*Time and space overhead-based comparison:*

In this section, we compare SPH and P-SPH in terms of memory and running time requirements. Here, we perform our analysis in two parts: (i) offline build phase where patterns are computed and a histogram is built during database statistics gathering time, and (ii) online phase, which involves the estimation of query predicate selectivity during query optimization time.

First, we present time and memory requirements for the offline build phase. Figures 20 and 21 show the build phase time and memory requirements, respectively, for both algorithms SPH and P-SPH.

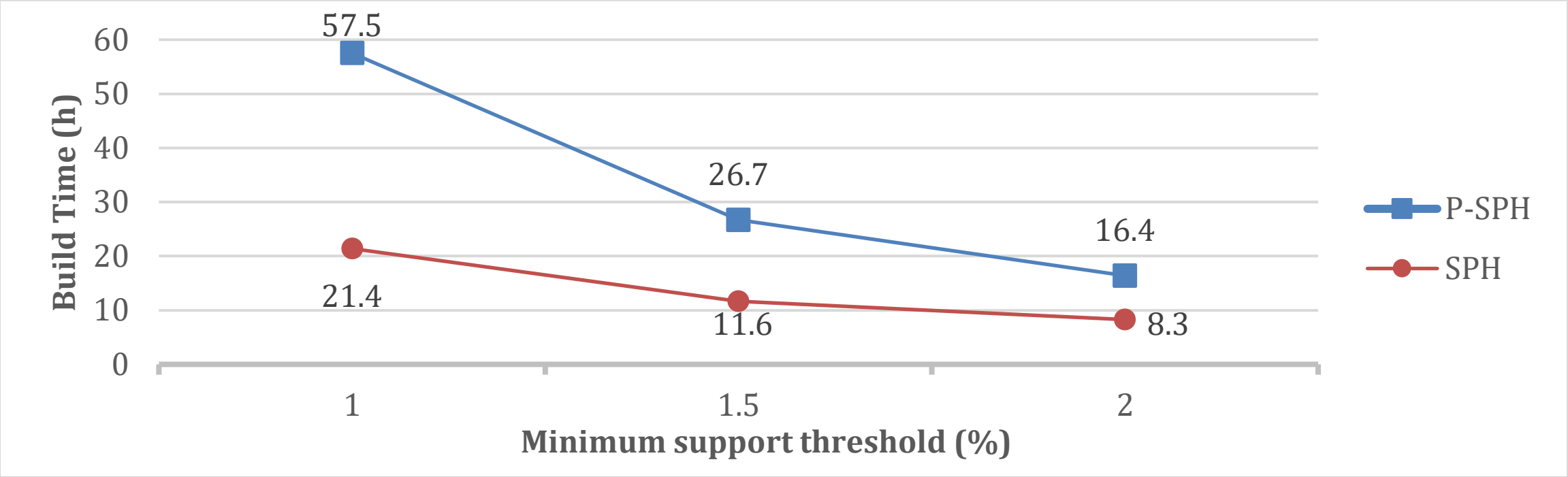

Figure 20: Average build phase time at different minimum support thresholds with 2048 buckets

***Observation 13:*** *In terms of build time, P-SPH takes considerably more time (around 2X at minsup = 2%) than SPH, and the time difference between approaches decrease, as the minsup values get higher.*

The above observation is directly explained by Figure 9, which shows that there is a similar relation between the numbers of positional and regular patterns. The number of positional patterns are significantly higher. Therefore, it takes more time to compute them. Besides, since this stage takes place offline before query time and possibly on a separate machine than the production servers, the time spent for this stage is considered a lot less critical than the online query optimization time, which is discussed, in the next set of experiments.

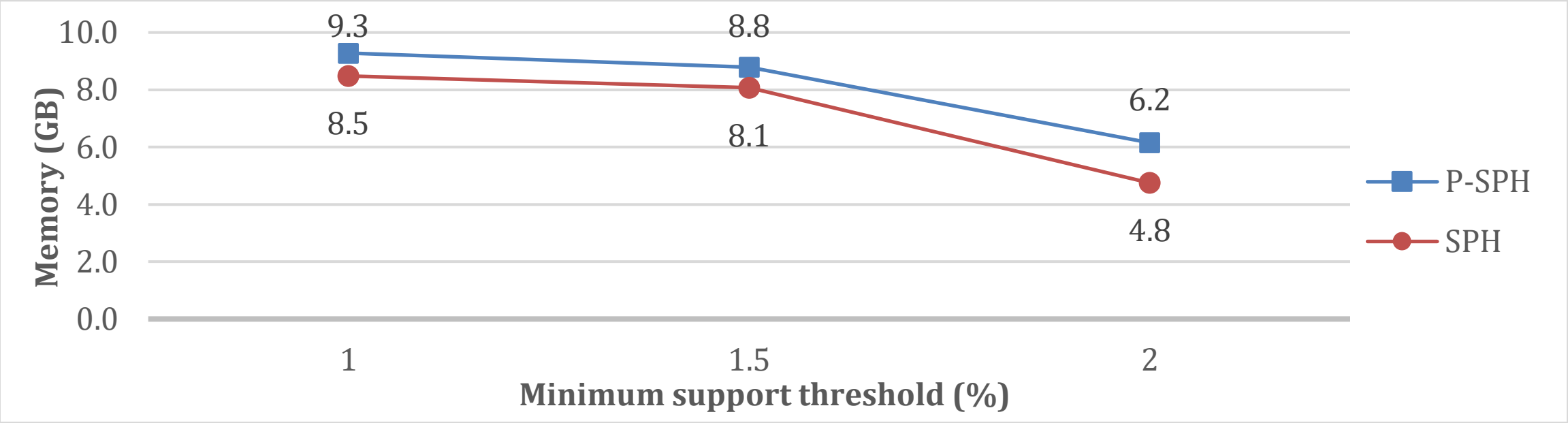

Figure 21: Average build phase memory overhead at different minimum support thresholds with 2048 buckets

***Observation 14:*** *In terms of build phase memory, P-SPH and SPH are comparable.*

Even though P-SPH computes 2 times more patterns that SPH, the memory requirements do not reflect this difference. This is mostly because the proposed positional sequence mining approach follows the depth-first strategy of the original BIDE algorithm that it extends.

Next, we present time and memory requirements for the online phase. Figure 22 shows the selectivity estimation time for different numbers of buckets at minimum support threshold of 1.5%. Then, Figure 23 presents the online phase memory requirements for both SPH and P-SPH.

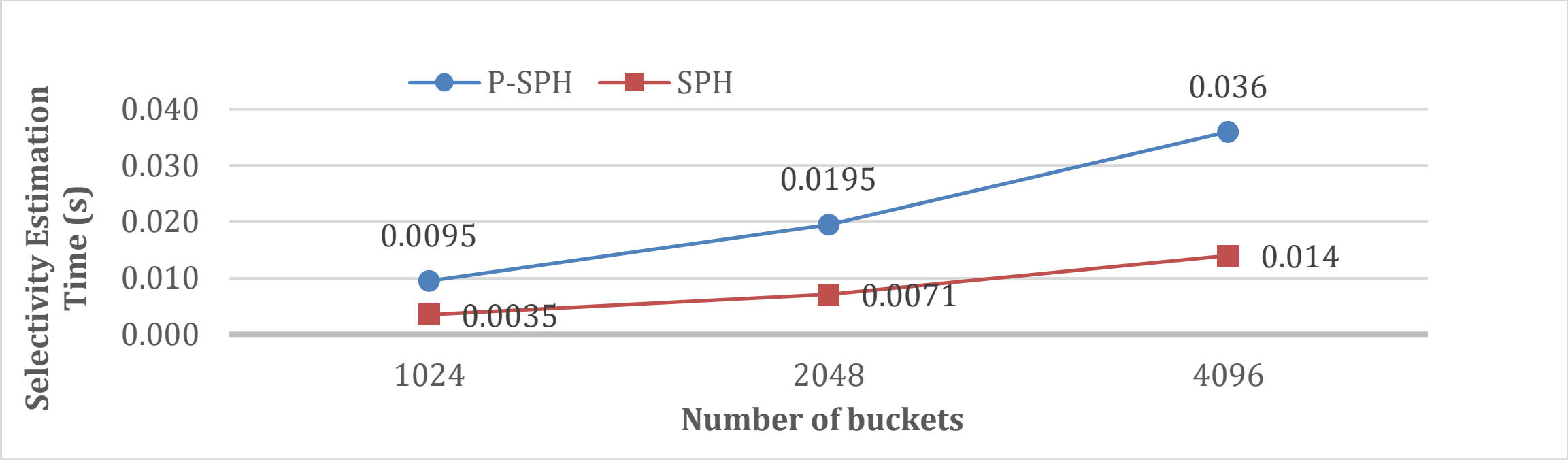

Figure 22: The average query time for all query types with varying number of buckets for P-SPH

***Observation 15:*** *The selectivity estimation time of SPH is at least 2 times less than that of P-SPH.*

There are two factors that contribute to the above observation: (i) P-SPH uses partitioning-based matching, which adds extra computation time, whereas SPH lacks this feature, and (ii) the endpoint values in histograms build with positional sequence patterns are longer than those built with regular sequence patterns. Hence, P-SPH spends more time than SPH while computing encapsulated or exact matches as well.

***Observation 16:*** *The query execution memory for the online phase is almost the same for both P-SPH and SPH.*

Both SPH and P-SPH store the same number of frequent closed patterns as endpoint values for the histogram. Since the histogram structure is the same, their online memory requirements are also similar with ignorable difference caused by a little longer patterns in the histograms for P-SPH.

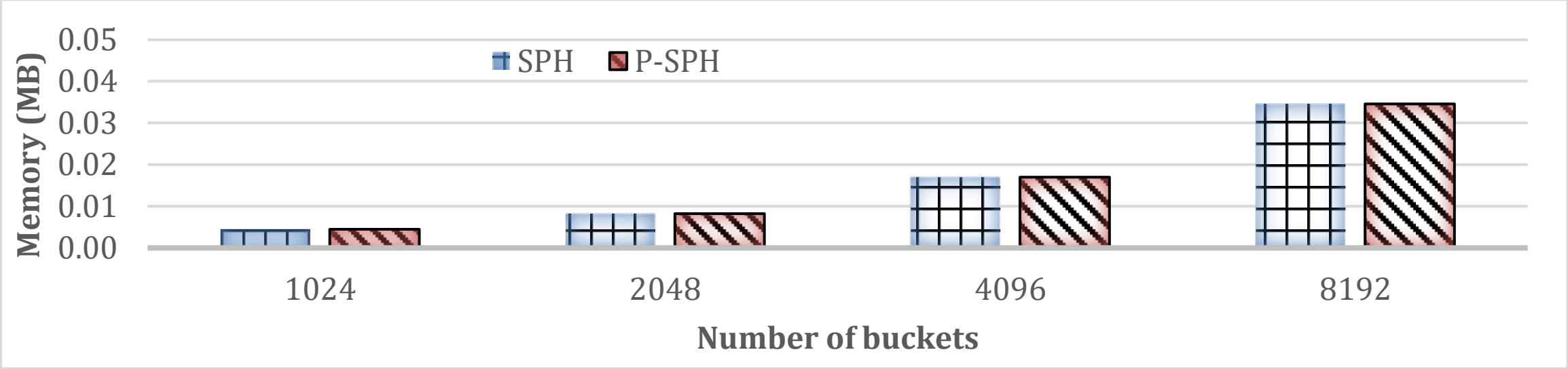

Figure 23: Online phase memory requirement comparison with varying number of buckets

*Accuracy-based comparison with earlier methods (KVI and LBS):*

In this section, we also compare P-SPH with two earlier methods, KVI (Krishnan et al., 1996) and LBS (Lee at al., 2009) in terms of the selectivity estimation accuracy. We also include SPH to illustrate all methods in one picture. Figure 24 shows the average relative error for the SPH, P-SPH, KVI, and LBS approaches.

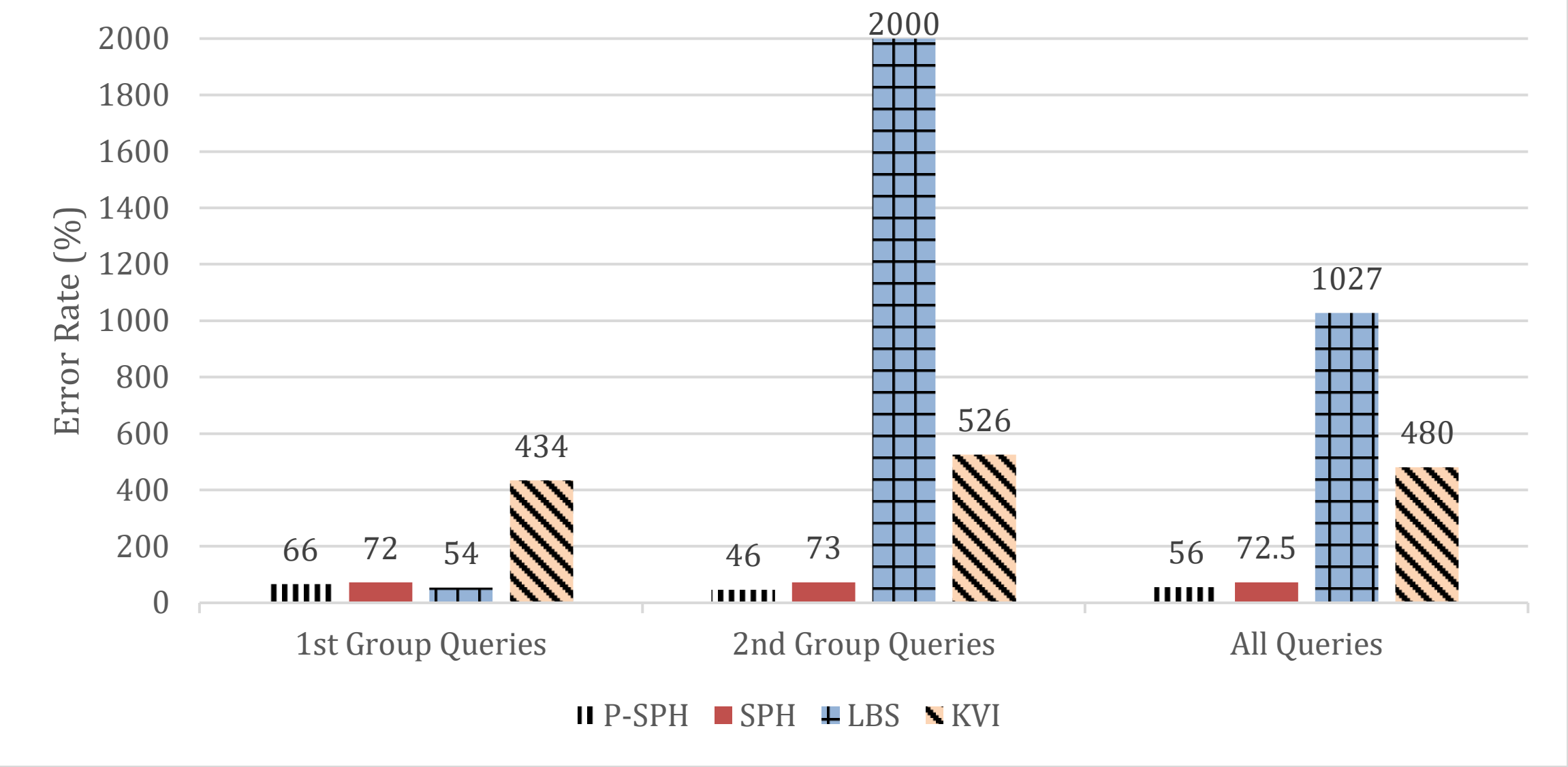

Figure 24: Accuracy comparison of SPH, P-SPH, KVI and LBS

***Observation 17:*** *P-SPH provides the least estimation error rate by a significant margin when the whole query set is considered, even though LBS performs slightly better for the first group of queries.*

## 7. Conclusion & Future Work

In this paper, we propose a new approach to estimate the selectivities of SQL LIKE query predicates. To this end, we introduce a new type of sequence patterns called positional sequence patterns, and extend a regular sequence mining algorithm to compute positional sequence patterns. A histogram is built on top of the mined positional sequence patterns during database statistics gathering time, and then this histogram is later employed during query optimization time to compute the estimated selectivities. In order to increase the coverage of histograms, we introduce an information content-based redundant pattern elimination approach. Besides, to take advantage of partial matches between histogram endpoints and query predicate strings, we also propose a partitioning-based matching algorithm. We assessed the proposed techniques on a real dataset from DBLP, and demonstrate that our methods significantly outperform the state of the art in terms of selectivity estimation accuracy.

As part of our future work, we plan to integrate our methods to a database management system, e.g., ORACLE, and evaluate different approaches to selectivity estimation in terms of their impact on query execution plans. To this end, we will employ industry standard benchmark query workloads, such as TPC-H.

## Acknowledgements

This research is in part funded by TUBITAK (Scientific and Technological Research Council of Turkey) under grant number 117E086.